**Single-molecule force spectroscopy with photoluminescent semiconducting polymers: Harnessing entropy**

Alessio Zaccone
Department of Chemical Engineering and Biotechnology
University of Cambridge
CB3 0AS Cambridge, U.K.
Email: az302@cam.ac.uk

*In this issue of Chem, Sprakel and co-workers use a fluorescence-doped flexible semi-conducting polymer to construct a single-molecule sensor which can detect ultra-weak forces in the molecular environment, with a grey scale down to 300 femtonewtons.*

When we take a rubber band and stretch it, the rubber band pulls back; releasing one hand will result in the rubber band taking back its original shape. This elastic, spring-like, response of a rubber, and all other polymeric materials that surround us in daily life, ranging from car tires, high-tech materials for airplanes and spacecraft to the protein structures that build the tissues in the human body, results from a particular property of the chain-like molecules from which these polymer solids are built. Each polymer chain within such a material has the shape of a single cooked strand of spaghetti; due to the thermal fluctuations that act at the molecular scale the chain can and will adopt a large diversity of different conformations even when the boundary conditions are fixed. The total amount of possible conformations that the chain can adopt results in something called conformational entropy. If we stretch the rubber band, we also stretch the spaghetti-like molecules inside the rubber band, which reduces the entropy of each molecule. As all systems tend to increase their entropy, this results in the molecules pulling back and resisting the stretching. The elasticity we observe in polymer materials thus results directly from entropy at the molecular scale. This is an example of entropic force which is of great importance to chemists, physical chemists and biochemists. In a completely different setting, and at much larger length scales, entropic forces have recently been invoked to explain the origin of the gravitational force, leading to the concept of "entropic gravity" [1].

For decades, the entropic elasticity of polymers could only be deduced indirectly, for example by measuring the temperature-dependence of the stiffness of a rubber band, which counter-intuitively increases with temperature due to its entropic origins. With the advent of the atomic force microscope and optical tweezers, the entropic nature of polymer elasticity, including bio-polymers like DNA [2] could be test directly for the first time, but only in very ideal conditions of dilute or semi-dilute polymers in a sea of solvent [3]. In real materials, however, where the polymer chains are surrounded by many peers, or in biological materials where the local surroundings are even more complex, only indirect proof exists for the effects of conformational entropy on the macroscopic response. Furthermore, it is very difficult to quantify the entropic force acting on a single chain inside such complex material, and the current physical understanding of deformation of bulk polymers relies on mean-field (coarse-grained) concepts such as the reptation tube introduced and studied by P.G. de Gennes and by S. Edwards and M. Doi in the 1970's-1980s'.

In their paper, Sprakel et al. have developed a creative way to solve this open challenge by taking advantage of the light-emitting properties of conjugated semiconducting polymers [4]. In the same way that a macroscopic spring of a coiled metal wire can be used to measure a force in the macroscopic world, the authors used a single polymer chain as an entropic spring to measure forces in the molecular realm. In the macroscopic analogue, we deduce a force by visually evaluating how far a spring has been extended and, when the spring constant of the spring is known, this extension can be converted into a force. In their experiments, the authors took this concept at the heart of old-fashioned Newtonian spring scales, and implemented it at the molecular scale for a polymer chain. To achieve a visual read-out of the extension of a single polymeric spring, whose size is nanometric and hence well below any resolution for direct visual observation, the authors made use of fluorescence energy transfer between different building blocks with the autofluorescent (semiconducting) polymers (see Fig. 1). This enabled them to measure the degree of extension of a single polymer chain embedded in a polymer-rich matrix using single-molecule fluorescence spectroscopy. The mechanism, depicted in Fig. 1., exploits the fact that, under optical stimulation, if the macromolecule is coiled, energy transfer between different non-adjacent segments of donor and acceptor monomers interspersed the chain is favoured over emission inside the individual donor/acceptor. Conversely, if the molecule is stretched, emission inside each monomer is favoured over energy transfer. This mechanism ingeniously provides a continuous spectrum from which it is possible to measure the state of the molecule (how stretched or coiled it is) and, from that information, extract the force acting on the macromolecule.

Because the response of a macromolecular spring is gradual and continuous, as it is for a macroscopic spring, they were able to demonstrate grey-scale deformation sensing at the molecular scale. When the deformations that are measured are converted into a force, using well-established models that describe the spring constant for an entropic spring, this leads to a new method to detect forces deep within materials at an unprecedented resolution of 300 femtonewtons, with a 100-fold improvement over existing fluorescence-based molecular sensors.

This new approach to resolve mechanical deformations at the macromolecular scale opens up a plethora of new possibilities to provide molecular insight into the complex macroscopic mechanics of polymers. In particular, the ability to perform greyscale measurements on a single chain, i.e. to not only tell that a molecule is deformed but to actually quantify how much this deformation is, is promising. When a polymer material, be it a rubber or a polymer glass (plastic), is mechanically deformed at the macroscopic scale, the question remains how this macroscopic deformation is distributed down to the scale of the individual chains and the segments (monomers) from which the material is made. Classical molecular elasticity models require the assumption that this redistribution is homogeneous (i.e. each molecular segment is displaced to a new position which is just the old position vector left-multiplied by the strain tensor), known as the assumption of *affine deformation*. However, over the last decades it has become very clear that in polymer materials this affinity assumption fails [5]. Owing to the disordered structure of the material at the molecular scale, force and strain distributions become inhomogeneous [6], where some chains in the material are deformed much more than the macroscopic deformation, while others may remain almost completely unaffected. This is known as the emergence of *nonaffinity* in the deformation [6] and is very important for solid polymers (e.g. glassy polymers [5]). It is a generic property of deformation of any glassy materials and differs somewhat from the Flory nonaffine model for rubbers due to the denser packing of molecules in the glass state. And even though these nonaffine effects are only noticeable at the molecular scale, and a macroscopic observer would remain completely unaware that they exist, it turns out that the mechanical properties of the material depend to a large extent on such nonaffine molecular effects. In the past decade, theoretical models [6] have revealed the importance of non-affinity, not only in glassy solids or random packings, but also to understand the puzzling mechanical properties of biological materials, e.g. actin filament networks [7], within our bodies. However, as there were no methods to measure the microscopic state of deformation of individual polymer chains within a material, these theoretical predictions could only be tested indirectly until now.

With the method established by Sprakel et al. the realm of molecular mechanics has now become tractable in relatively straightforward experiments. This should allow a new step to be taken in linking molecular mechanics to macroscopic properties in which these non-idealities are explicitly accounted for. In fact, in their paper, the authors demonstrate the ability to quantify nonaffinity by comparing the state of deformation of hundreds of individual chains within the same matrix that has been deformed. And low-and-behold, the nonaffinity that was predicted by theory could be directly observed, even showing a quantitative agreement between the measurements and theoretical predictions from the literature [8,9].

While simple in concept - using the light-emitting properties of a doped semiconducting polymer, to generate a mechanical spectrum at the molecule scale - the implications of this methodological development are broad. Besides the elucidation of mechanical response and polymer dynamics at the molecular scale, another application could be the identification of the pathways taken by charges in semiconducting polymer materials for solar cells, where the charge mobility is linked in a non-trivial way to the macromolecule conformation [10].

However, challenges, both in the implementation of this method and the method itself remain; for example, chemical inhomogeneity in the polymeric force sensors, leads to signal broadening. Improvement of the synthetic protocols would enable a more fine-tuned detection of forces in which the artificial signal due to chemical heterogeneity is suppressed. Moreover, a full mapping of local deformations within a three-dimensional solid requires extension of the method to spectrally-resolved confocal microscopy which enables high resolution observation in full three dimensions. One could even envision some extension of the current method to beat the diffraction limit and go towards super-resolved force imaging, through the clever design of photo-switchable force sensors that can be addressed individually.

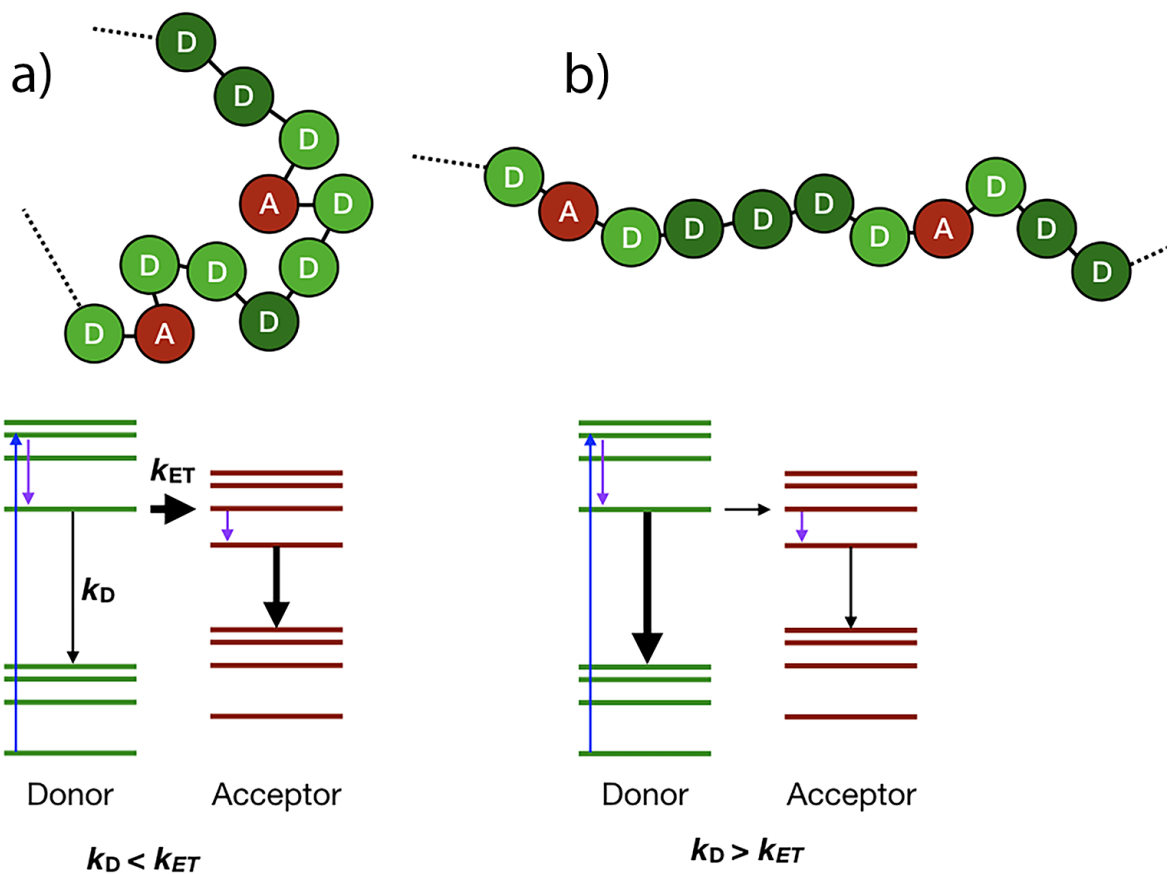

**Figure caption.**

A conductive conjugated polymer chain is doped with acceptor and donor monomers and stimulated with light. Transition to a higher energy level of a donor monomer is excited by the applied radiation. If the donor is surrounded by other donors only, the excited state decays via an internal photon emission process within the donor monomer, with rate $k_D$. When an acceptor monomer is sufficiently close to a donor monomer, instead, the excitation energy can be transferred from the donor to the acceptor which then emits a photon, via Förster resonant energy transfer, with rate $k_{ET}$. The rate of the donor decay $k_D$ is fixed, but the rate of energy transfer $k_{ET}$ depends on the distance and number of acceptor-donor pairs. The black arrow weight indicates the preference for either donor emission or energy transfer and subsequent acceptor emission, and thus codes the relative ratios of donor to acceptor light that can be observed. For the coiled (unstretched) chain in panel (a), the local monomer density is high and thus $k_{ET} > k_D$; conversely, for the stretched chain in panel (b) the local monomer density is low such that most emission occurs by direct decay of the donor. The purple lines are the thermal decay processes, since the excited state energy is always higher than the energy of the emitted photon (the emission is redshifted with respect to the absorption). In (a) the chain is more densely coiled, whereas in (b) it is more stretched: it follows that the chain conformation on the left is in a higher entropy state compared to the chain in (b). The chain in (b) is therefore subjected to a strong entropic force. From the measured emitted light spectrum and intensity it is therefore possible to infer back the mechanical force acting on the chain.